\documentclass[journal,twocolumn]{IEEEtran}
\ifCLASSINFOpdf
\else
\fi
\usepackage{amssymb}
\usepackage[figuresright]{rotating}

\usepackage{subfigure}
\usepackage{float}
\usepackage{pgfplots}
\usepackage{amsmath}
\newtheorem{theorem}{Theorem}
\newtheorem{lemma}{Lemma}
\newtheorem{definition}{Deifinition}

\usepackage{algorithmic}

\usepackage{array}

\usepackage{color}
\usepackage{fixltx2e}
\usepackage{stfloats}
\usepackage{url}
\usepackage{enumerate}
\usepackage{multirow}
\usepackage{multicol}
\usepackage{bm}
\usepackage{indentfirst}
\usepackage{booktabs}
\usepackage{tikz}
\usepackage[usenames,dvipsnames]{pstricks}
\usepackage{epsfig}
\usepackage{pst-grad} 
\usepackage{pst-plot} 
\usepackage[space]{grffile} 
\usepackage{etoolbox} 
\usepackage[justification=centering]{caption}
\begin{document}
%
\title
{Privacy-Assured Outsourcing of Compressed Sensing Reconstruction Service in Cloud}
%
%
%
\author{Ping Wang
\thanks{P. Wang was with the College of Electronics and Information Engineering, Southwest University, Chongqing 400715, China (e-mail: bruce\_wp@163.com).}
}
\maketitle

\begin{abstract}
Compressed sensing (CS), breaking the constriction of Shannon-Nyquist sampling theorem, is a very promising data acquisition technique in the era of multimedia big data. However, the high complexity of CS reconstruction algorithm is a big trouble for endusers who are hardly provided with great computing power. The combination of CS and cloud has the potential of freeing endusers from the resource constraint by cleverly transforming computational workload from the local cilent to the cloud platform. As a result, the low-complexity encoding virtue of CS is fully leveraged in the resource-constrained sensing devices but its highcomplexity decoding problem is effectively addressed in cloud. It seems to be perfect but privacy and security concerns are ignored. In this paper, a secure outsourcing scheme for CS reconstruction service is proposed. Experimental results and security analyses demonstrate that the proposed scheme can restrict malicious access, verify the integrity of the recovered data, and resist brute-force attack, ciphertext-only attack, and plaintext attack.
\end{abstract}


%
\IEEEpeerreviewmaketitle

\section{Introduction}
%
%
%
%

\IEEEPARstart{W}{ith} the development of multimeadia big data, how to acquire data efficiently is an important topic all the time.
In the traditinal infromation acquisition system, the encoder usually requires to acquire massive amounts of data following Shannon-Nyquist sampling theorem and then compress it for efficient usage of storage and bandwidth resources. Such a sampling-then-compression framework works in a resource-unfriendly manner and would take a great deal of complexity on the design of sensing system. Compressed sensing (CS) is a newly-developing data acquisition and reconstruction technique mixing the signal sampling and data compression processes by virtue of the sparsity of signal to be captured. 
In the CS paradigm, the combination of sampling and compression could not only save a considerable lot of computing and storage resources but also cut down the hardware overhead sharply. In the past decade, CS takes a significant step from theory to practice, which is widely used in magnetic resonance imaging\cite{lustig2008compressed}, sparse channel estimation\cite{berger2010application}, and internet of thighs (IoT)\cite{li2012compressed}, etc. All of these take advantage of the low-complexity encoding virtue of CS. 

Every coin has two sides, and CS is no exception. In the CS-based applications, the low-complexity encoding is at the cost of the high-complexity decoding. Accordingly, CS can be called as a magic technique of shifting the complexity from the encoder to the decoder. In essence, decoding the captured data, namely sparse signal reconstruction (SSR), is to solve a linear programming problem which requires a substantial amount of computational power. For the resource-limited endusers, it is not an advisable choice to put such a heavy workload on the terminal devices, such as mobile phone and some wearable IoT devices. On the one hand, the workload of SSR task could not usually be well-matched with the computational power of the terminal devices. On the other hand, performing SSR in the local client is a time-consuming task. Therefore, how to perform the SSR task effectively has always been a key challenge of restricting the CS-based applications. With the development of cloud computing, an alternative approach is to outsourced the SSR task to a cloud.

Cloud Computing permits convenient on-demand network accesses to a shared pool of configurable resources (e.g., networks, storage, and computations) that can be rapidly provided and released with great efficiency and minimal management effort[]. Delivering the abundant resources to the customers is the main service of cloud providers. By outsourcing the workloads to the cloud, the computational power of the endusers is no longer limited by their resource-constrained devices. Through the above discussions, outsourcing SSR services to a cloud seems to be feasible and promising. However, there still exist a number of fundamental and critical challenges, among which security is the top concern.

With the combination of CS and IoT, the outsourced SSR service often contain some sensitive information, such as military radar data and personal health monitoring data. Most traditional cryptosystems, providing a high level of security guarantee at the cost of a heavy computation workload, is hardly embedded in the physical layer of the lightweight sensing applications. Therefore, there exist some practical security challenges in such an outsourced framework where the high-complexity SSR task is transformed from the resource-constrained sensing devices to the cloud and the reconstructed data are sent to the legitimate enduers. At  first, how to guarantee the security of transmitting sensitive data from the sensing devices to the cloud servers is of great concern in the absence of specialized cryptosystems. Secondly, the reconstructed data should be secret for the semi-trusted clouds but available for the legitimate endusers. In addition, from the perspective of a cloud provider, it plays a pivotal role how to check the legality of enduers asking for an access to the reconstructed data. Finally, how to avoid the tampering attack to the reconstructed data but permit content-preserving data distortions is a big challenge all the time, resulting from that the reconstructed data are doomed to be different from the original data to some extent.

In this paper, a privacy-assured cloud-assisted outsourcing scheme is proposed for the SSR service. Without any additional cryptosystem, two privacy-preserving primitives, compressed sensing-based encryption and sparse represent-based encryption, are seamlessly embedded in the process of CS encoding. The cascade effect of two primitives provides a high level of transmission secrecy guarantee for the captured data. Sparse represent-based encryption makes sure that the reconstructed data would not leak any information except for sparsity. To restricting malicious access, a kind of access passwords is extracted from some prior encoding knowledge and then regarded as the certificate of authorized endusers. Besides, CS-based message authentication codes are generated to figure out whether the receiving data have been pulluted or tampered. 

 The rest of this paper is organized as follows. Some related works are briefly introduced and analyzed in Section 2.  Section 3 gives the basic knowledge of CS, system architecture, threat model, and design goals. In Section 4, several basic techniques are proposed, including bi-level encryption, integrity verification, and access control. Section 5 describes our privacy-assured outsourcing scheme in detail. Experimental results and analysis are given in Section 6 . A conclusion is drawn in Section 7. 

\section{Related works}

SSR is essentially a linear programing problem and its solution generally needs the support of a great computational power. Outsourcing the high-complexity CS decoding task to the cloud is a newly well-studied flied. Secure computation outsourcing\cite{atallah2002secure,hohenberger2005securely,atallah2010securely} is to protect the input and ouput privacy in the process of outsourced computation. As a breakthrough work, the fully homomorphic encryption (FHE) was proposed to enables directly performing functions on encrypted data while yielding the same encrypted results as if the functions were run on plaintext. Aftherwards, some secure outsourcing schemes based on FHE were proposed\cite{kerschbaum2012outsourced,article,li2018privacy}. However, they are usually impractical because of the exceedingly high complexity of FHE operation. Note that SSR workloads mainly consist of matrix transformation and linear equation computations. Lei et al. proposed a number
of secure outsourcing protocols for solving matrix inversion\cite{lei2013outsourcing}, matrix multiplication\cite{lei2014achieving}, and matrix determinant computations\cite{lei2014cloud}. Wang et al. presented some privacy-preserving outsourcing protocols for linear equation\cite{5961757,6231624} and liner programming\cite{5935305,7070751} in cloud. Besides, Wang et al. further proposed a secure outsourcing scheme for CS-based image reconstruction service\cite{6562794} but its complexity is still too high for the resource-constrained devices To achieve secure SSR outsourcing as possible in a low-complexity manner possible, some newly works tried to transform SSR problem to the encrypted one in the process of SR\cite{zhang2015support,hu2017compressive,zhang2019efficiently}. All of those aimed to embed cryptographic operation in the process of spare representation (SR). Such an ideology occurred in some CS-based privacy-preserving works\cite{zhang2016bi,zhang2017novel,zhou2018double}. However, SR-based encryption is a weak symmetric cipher such that anyone holding abundant computing resources can reveal the sparsity of the plaintext. It remain challenging to provide outsourced data with high level of security guarantee in a resource-constrained scenario.

\section{Problem formulation} 
		This section introduces the basic knowledge of CS, system architecture, threat model, and our design goals.

\subsection{CS preliminaries}
According to CS theory \cite{Donoho2006Compressed, 4472240, 1580791}, there are two preconditions of guaranteeing the successful reconstruction from very few samples. One is that the signal to be sampled should be sparse or could be sparsely represented in a certain basis. Fortunately, most natural signals are sparse under an appropriate transformation domain. The other is that the measurement matrix should be as incoherent with the basis matrix as possible. 
CS theory can be mathematically summarized as follows.
 An one-dimensionality signal ${\bf x} \in \mathbb{R}{^N}$ is sparsely represented as a coefficient vector ${\bf s} \in \mathbb{R}{^N}$ containing at most $K$ non-zero elements in a certain basis ${\bf\Psi}  \in {\mathbb{R}^{N \times N}}$ as follows:
\begin{equation}
\bf x = { \bf \Psi } s, 
\end{equation}
where ${\left\| {\bf s}  \right\|_0} = K \ll N$ and ${\left\| {\cdot}  \right\|_0} $ counts the number of the non-zero entries. In the CS paradigm, such a coefficient vector ${\bf s}$ can be sampled and compressed simultaneously by the sensing matrix ${\bf A} \in {\mathbb{R}^{M \times N}}$ as follows:
\begin{equation}
{\bf y}  = {\bf A  \Psi }^{ - 1} {\bf x}  = {\bf A s},
\end{equation}
where ${\bf y} \in \mathbb{R}{^M}$ $(K < M \ll N)$ denotes a measurement vector (i.e., sampled data) and ${\bf \Phi} ={\bf A}{\bf \Phi}^{-1} $ denotes the measurement matrix. ${\bf A} $ must obey the energy-preserving guarantees \cite{Candes2006Near, Baraniuk2008, KUENG2014110}.

Evidently, the linear systems $(2)$ does not have the unique solution. Considering the sparsity of this signal, the intuitive way is to solve ${l_0}$ norm problem known as greedy iteration with denoising as follows:
\begin{equation}
{\bf s}'  =  \mathop {\arg \min }\limits_{\bf s}  {\left\| \bf s  \right\|_0}   \quad s.t. \quad{\rm{ }}{\left\| {\bf y -  A s } \right\|_2} \le \varepsilon, 
\end{equation}
where $\varepsilon$ indicates noise levels. In essence, solving $(3)$ is a NP-hard problem because it works with exhaustively searching over all column subsets of $\bf{ A}$, which can be relaxed as ${l_1}$ norm problem  \cite{Candes2006Near} known as basis pursuit with denoising as follows:
\begin{equation}
{\bf s}'  = \mathop {\arg \min }\limits_{\bf s}  {\left\| \bf s  \right\|_1}\quad s.t. \quad {\left\| {\bf y - A s } \right\|_2} \le \varepsilon,
\end{equation}
where ${\left\| {\cdot}  \right\|_1} $ counts the sum of absolute value of all elements. The basis pursuit algorithm based on solving $(4)$ is a linear programming problem, thus it can relatively accurately reconstruct this original signal at the cost of high computation. 

Except for the sparsity, there is another precondition that ${\bf \Phi}$ should be as incoherent as possible with ${\bf \Psi}$. Strictly speaking, the incoherence means that any basis vector in ${\bf \Psi}$ can not be represented in the basis composed of the rows of ${\bf \Phi}$. To meet such a incoherence requirement, a mathematical definition of Restricted Isometry Property (RIP) was given in \cite{Baraniuk2008}, in which a matrix ${\bf  \Phi}$ is said to satisfy the RIP of order $K$ if there exists a ${\delta_K} \in (0,1)$ such that
\begin{equation}
(1 - {\delta _K})\left\| \bf s \right\|_2^2 \le \left\| {{\bf  \Phi} \bf s} \right\|_2^2 \le (1 + {\delta _K})\left\| \bf s \right\|_2^2
\end{equation}
holds for all $k$-sparse signal $\bf s$.\\

\subsection{System architecture and threat models}
\begin{figure}[t]
	\centering
	\includegraphics[scale=0.45]{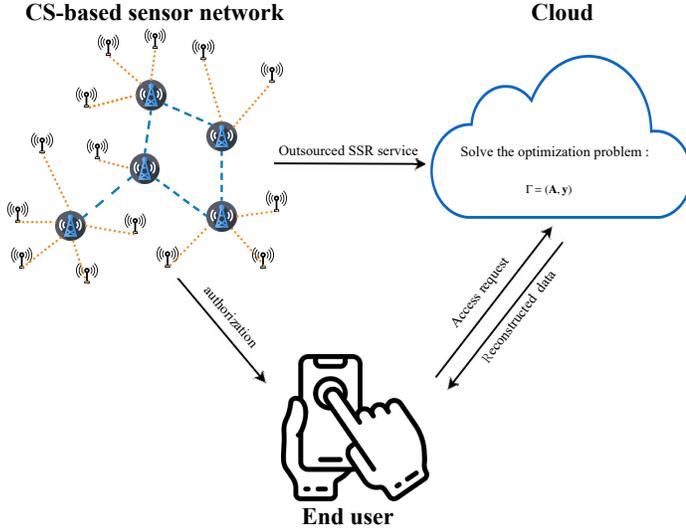} 
	\caption{System architecture.} 
	\label{fig1}
\end{figure}
By virtue of low-complexity encoding, CS has a great potential of being considered as a low-overhead sensing technique widely used in the physical layer of IoT. With the combination of cloud and IoT, a promising CS-based application scenario is shown in Fig. \ref{fig1}. In a CS-based sensor network, the subnodes capture the original data $\bf x$ in a cost-effective manner and then the parent nodes outsource the SSR task ${\Gamma}= ({\bf A},{\bf y})$ to a remote cloud. Afther receiving the access request from the enduser, the cloud manager firstly verify the legitimacy of this enduser and then determine whether or not to send the reconstructed result. In the local side, the enduser firstly figure out whether the received data ${\bf s}'$ haven been tampered or polluted and then gets the final data ${\bf x}'$ by a simple computation ${\bf x}' = {\bf \Psi}{\bf s}'$ .

In this paper, we consider a semi-trusted cloud. Specifically, the cloud provider is supposed to honestly perform SSR task but very curious about the information contained by computational results.
There exist three main threats in the above system architecture, resulting from two deep-rooted practical issues. One is how to provide the caputured data with high confidentiality in the resource-constrained sensing player. The other is how to buid the trust mechanism in the process of sensor-cloud-user interaction. For each sensor node, most traditional high-energy cryptosystems are unpractical such that the captured privacy-sensitive data are attractive to the malicious attackers. In the absence of the privacy-assured measures, senor and user lose the ownship of the reconstructed result as a result of outsourcing SSR task to the cloud. In the process of cloud-user interaction, attackers may disguise themselves as a legitimate enduser to access to the reconstructed result. In addtion, the atttackers may tamper the sampling data to make SSR fail or distort original information.
\subsection{Design Goals }
In consideration of the privious system architecture and threat models, we proposed a privacy-assured outsourcing framework for SSR service in cloud to meet the following design goals:
\begin{itemize}\setlength{\itemsep}{1pt}
	\item Transmission confidentiality: The confidentiality of the captured data is cleverly guaranteed without a specialized cryptosystem.
	\item Privacy-assured outsourcing: The semi-trusted cloud has to honestly perform the SSR task and has no ability of stealing the privacy contained in the reconstructed data.
	\item Access control: The malicious access to the reconstructed data can be efficiently restricted.
	\item Integrity verification: The energy-infinite noises can be tolerated but the tampering attack can be exactly detected.
\end{itemize}	

\section{Basic techniques} 
	In this section, we propose three key techniques, including bi-level encryption, integrity verification, and access control.

\subsection{Bi-level encryption}
Most existing high-complexity cryptosystems usually could not be matched with the resource-constrianed information acquisition system such that how to guarantee the confidentiality of the captured data are an ubiquitous challenge. In this paper, we propose a bi-level encryption scheme based on CS, where the cryptographic features are embedded in both subsampling and sparse representation processes. Specifically, two key-controlled sensing matrix ${\bf \Phi}_k$ and basis matrix ${\bf \Psi}_k$ are designed, renewed, and kept secret.

In \cite{zhang2016bi},  Zhang at el. indicated that andomly exchanging the position of rows and multiplying rows by a non zero number would not alter the SR competence of a certain sparsifying basis, which can be mathematically summarized in Theorem 1. 
\begin{theorem}
	Two basis matrices ${\bf \Psi}$, ${\bf \Psi}'$ are said to be equivalent, namely ${\bf x} = {\bf \Psi s} = {\bf  \Psi}' {\bf s}'$, ${\left\| {\bf s} \right\|_2}={\left\| {\bf s}' \right\|_2} $, if given any permutation matrix ${\bf P}$ and diagonal matrix ${\bf D}$ there exists ${\bf \Psi}' = {\bf \Psi} {\bf P D}$.
\end{theorem}
Although the spare representations are different under different sparsifying basis, their sparsity would be doomed to be same. According to it, the sparsity representation process can be viewed as a symmetric encryption process as follows:
\begin{equation}
{\bf s} = ({{\bf \Psi} }{{\bf P}_{k_1}}{{\bf D}_{k_2}})^{ - 1}{\bf x} = {{\bf \Psi} ^{ - 1}_{k}}{\bf x}
\end{equation}
where ${\bf x}$, ${\bf s}$, and $ {{\bf \Psi} _{k}} = {{\bf \Psi} }{{\bf P}_{k_1}}{{\bf D}_{k_2}}$ are regarded as the plaintext, the ciphertext, and the secret, respectively, as well as ${\bf P}_{k_1}$ and ${\bf D}_{k_2}$ denote the key-controlled random permutation matrix and diagonal matrix, respectively. Accordingly, the semi-trusted cloud providers could not get the privacy-sensitive ${\bf x}$ from the reconstructed ${\bf s}$ with the absence of ${{\bf \Psi} _{k}}$.

Some previous works encrypts and sparsify the original signal simultaneously and then directly outsource the finally acquired data ${\bf y} = {\bf A} {\bf s}$ to the cloud\cite{zhang2015support,hu2017compressive,zhang2019efficiently}. In their outsourcing framework, the sparsity of the plaintext is a public information for everybody with strong computational power. From the perspective of cryptography, their outsourcing frameworkes are vulnerable and fragile. To fix such a security flaw, we try to embed the cryptographic features in the subsampling process, namely CS-based encryption. 

CS-based encryption is a recently proposed lightweight cipher, in which ${\bf s}$, ${\bf y}$, and ${\bf A}$ are regarded as the plaintext, the ciphertext, and the secret key, respectively, as well as the subsampling and the reconstruction processes as the encryption and the decryption processes, respectively. The key is to make sure that the sensing matrix ${\bf A}$ is known by the encoder and the decoder and satisfies the RIP condition. Considering the security and efficiency, we employ the structurally subsampled matrix to construct ${\bf A}$ controlled by several parameters and satisfying RIP condition.

At fist, a block identity matrix ${\bf S}  \in {\mathbb{R}^{M \times N}}$ is constructed as defined in Definition 1, and a diagonal matrix ${\bf R}_{k_3}  \in {\mathbb{R}^{N \times N}}$ whose non zero elements are independent symmetric Bernoulli or Rademacher random variables is generated and renewed by changing the original conditions or inputs of pseudo random number generator (PRNG). Then, a structurally subsampled matrix is generated as follows:
\begin{equation}
{\bf A}_{k} = {\bf S}{\bf R}_{k_3} {\bf \Psi},
\end{equation}
where $\bf \Psi$ denotes a unitary matrix (i.e., sparsifying basis). Essentially, the combination form ${\bf S}{\bf R}_{k_3}$ is a row-mixing matrix. According to Theorem 2 given in \cite{bajwa2009restricted}, such a ${\bf A}_{k}$ has been proved to be a RIP matrix.
\begin{definition}
	Given $N$ and its divisor $L$, a block identity matrix is defined as ${\bf S}=[{{S}_1^T}, {{S}_2^T},..., {{S}_M^T} ]^T$, whose $i$-th row vector $[{{\bf 0}^{1 \times (i - 1)L}}{{\bf 1}^{1 \times L}}{{\bf 0}^{1 \times (N - iL)}}]$ containing $L$ adjacent $1$s is $L$-sparse, $i \in [1,M]$, $M = {N \mathord{\left/
			{\vphantom {N L}} \right.
			\kern-\nulldelimiterspace} L}$.
\end{definition}
\begin{theorem}
\cite{bajwa2009restricted}	Let ${\bf A} \in {\mathbb{R}^{H \times N}}$ be a structurally subsampled matrix, randomly subsampled from the unitary matrix $\bf U$ and the row-mixing matrix ${\bf S}{\bf R}$ and then normalized column by column, namely, ${\bf A} = \sqrt {\frac{M}{H}} {\bf S}{\bf R}{\bf U}$, $H \le M$. For each integer $a>2$, any $b>1$, and any $\delta_K  \in ({\rm{0}},{\rm{1}})$, there exist two absolute positive constants $c_1$ and $c_2$ such that the matrix ${\bf A}$ satisfies the RIP of order $K$ with probability at least ${\rm{1 - 20max\{ exp( - }}{{\rm{c}}_2}\delta _K^2{\rm{z),}}{{\rm{N}}^{ - 1}}{\rm{\} }}$ if
	\begin{equation}
{\rm{H}} \ge {c_1}zKN{\mu ^2}({\bf U} ){\log ^3}N{\log ^2}K,
\end{equation}
where ${\mu}( {\bf U} )$ denotes the coherence coefficient of ${\bf U}$, namely
\begin{equation}
\mu ({\bf U} ) = \mathop {\max }\limits_{1 \le i \ne j \le N} \frac{{\left| {\left\langle {{\psi _i},{\psi _j}} \right\rangle } \right|}}{{{{\left\| {{\psi _i}} \right\|}_2}{{\left\| {{\psi _j}} \right\|}_2}}}.
\end{equation}
\end{theorem}
Unifying SR and subsampling processes, the CS projection can be expressed as :
\begin{equation}
{\bf y} =  {\bf S}{\bf R}_{k_3} {\bf \Psi} ({{\bf \Psi} }{{\bf P}_{k_1}}{{\bf D}_{k_2}})^{ - 1}{\bf x} = {\bf A}_{k} {\bf s}.
\end{equation}
The above format implies that the original signal $\bf x$ is sparsely represented in the transformation domain ${\bf \Psi}_k$ and the sensing matrix ${\bf A}_{k}= {\bf S}{\bf R}_k {\bf \Psi}$ is generated from the row-mixing matrix ${\bf S}{\bf R}_k$ and ${\bf \Psi}$ in a structurally subsampling manner. Although there exist some differences between ${\bf \Psi}$ and ${\bf \Psi}_k$, ${\bf A}_{k}$ is still a RIP matrix, which can be inferred from Theorem 3. 
\begin{lemma}
	For two equivalent basis matrix ${\bf \Psi}$ and ${\bf \Psi}'$, their coherence coefficients are also equivalent, namely $\mu ({\bf \Psi} ) =\mu ({\bf \Psi}' ) = \mu ({\bf \Psi}, {\bf \Psi}')$.
\end{lemma}
{\bf Proof.}  The equivalence between ${\bf \Psi}$ and ${\bf \Psi}'$ means that there must be a permutation matrix $\bf P$ and a diagonal matrix $\bf D$, which satisfy ${\bf \Psi}' = {\bf \Psi} {\bf P D}$. Aiming at column permutations, $\bf P$ obviously does not affect the coherence. $\bf D$ aims to multiply some rows by non zero numbers. It makes the inner product and 2-norm increase by equal times such that the coherence still retains unchanged. All in all, ${\bf \Psi}$ and ${\bf \Psi}'$ do not the coherence and the above lemma holds.
\begin{theorem}
	If a structurally subsampled matrix ${\bf A}$, generated from an unitary matrix $\bf U$ and a row-mixing matrix ${\bf S}{\bf R}$, satisfies the RIP with high probability, a matrix ${\bf A}'$ satisfying RIP with equally high probability can be generated from ${\bf U}'= {\bf U} {\bf P} {\bf D}$ and ${\bf S}{\bf R}$, where ${\bf P}$ and $ {\bf D}$ denote the permutation matrix and the diagonal matrix, respectively.
\end{theorem}
{\bf Proof.} The above theorem implies that permutation matrix and diagonal matrix do not affect the RIP condition for any structurally subsampled matrix. 
$1$) ${\bf A}$ is generated by subsampling ${\bf SR} {\bf U}$ and then normalizing its column vectors. The normalize operation makes ${\bf D}$ in vain. So subsampling ${\bf U}{\bf P}{\bf D}$ is equivalent to subsampling ${\bf U}{\bf P}$. Furthermore, ${\bf U}{\bf P}$ is also an unitary matrix.  

\noindent According to Theorem 2, ${\bf A}'$ satisfy RIP of order $K$ with probability at least ${\rm{1 - 20max\{ exp( - }}{{\rm{c}}_2}\delta _K^2{\rm{z),}}{{\rm{N}}^{ - 1}}{\rm{\} }}$ if
	\begin{equation}
{\rm{H}} \ge {c_1}zKN{\mu ^2}({\bf U}{\bf P}{\bf D} ){\log ^3}N{\log ^2}K.
	\end{equation}
According to Lemma 1, ${\mu}({\bf U}{\bf P}{\bf D}) = {\mu}({\bf U})$ ensures that $(11)$ is equivalent to $(9)$. So the above theorem holds.

By keeping the keys $(k_1, k_2, k_3)$ secretly renewed between the encoder and the decoder, such a projection like $(10)$ is a bi-level symmetric encryption process. $(k_1, k_2)$ are regarded as the secret keys of SR-based encryption and $ k_3$ as the secret key of CS-based encryption.

\subsection{Integrity verification}
Outsourced computations involve multi individuals, including sensor, cloud, and enduser. Due to the presenceof maliciou attackers and the absence of trust mechanism, data may be polluted or tampered in the process of transmission, storage, and computing. So guaranteeing data integrity is of great importance. Unfortunately, CS data integrity verification is a difficult work in the resource-limited scenarios. One is how to generate message authentication codes $MAC$ directly for a analog signal. The other is how to distinguish malicious tamper and acceptable noise in view of  the robustness of CS measurements.

It is well known that CS is a lossy and lightweight data acquisition technique such that most existing integrity verification algorithms like MD5 and SHA2, which are very sensitive for data perturbation and endowed with high complexity, are infeasible in the CS-based applications. Therefore, it is challenging to verify the integrity of the received data in the CS scenarios.

By leveraging the compressive feature and breaking RIP condition, an one-way hash matrix ${\bf A}_{MAC} \in {\mathbb{R}^{m \times N}}$ $(m<<N)$ can be constructed to get the authentication measurements ${\bf y}_{MAC}$ of sparse coefficients $\bf s$ while subsampling as follows:
\begin{equation}
{{\bf y}_{MAC}} = {{\bf A}_{MAC}}{{\bf \Psi} ^{ - 1}}{\bf x} = {{\bf A}_{MAC}}{\bf s}.
\end{equation}
Note that the one-way guarantee of the above process depends on the non-RIP condition of ${\bf A}_{MAC}$.  Considering that, we suggest that ${\bf A}_{MAC} = {\bf 
\Phi}_{MAC}{\bf \Psi}$ is a non-RIP matrix by making ${\bf 
\Phi}_{MAC}$ as incoherent with ${\bf \Psi}$ as possible and letting $m$ less than what is required by the RIP condition.

As studied in \cite{duarte2011structured}, the relation between the incoherence and the number of the required measurements is given in Theorem 4. Accordingly, it is not hard to draw a conclusion that the matrix ${\bf A}_{MAC}$ is one-way as long as its rows number is less than $CK\sqrt N t\mu ({\Phi ^T},\Psi )\log (tK\log N){\log ^2}K$. 
\begin{theorem}
 \cite{duarte2011structured} For a fixed basis matrix $\bf \Psi$, the matrix $\bf \Phi \Psi \in {\mathbb{R}^{M \times N}}$ is said to satisfy the RIP condition with probability as least $1 - 5{e^{ - t}}$ if 
\begin{equation}
M \ge CK\sqrt N t\mu ({\Phi ^T},\Psi )\log (tK\log N){\log ^2}K
\end{equation}
\end{theorem}

To relieve the transmission burden and increase the entropy, $MAC$ are generated by extracting the sign of ${{\bf y}_{MAC}}$ and then performing non-return-to-zero (NRZ) line code as follow:
\begin{equation}
MAC = {f_{NRZ}}(sign({{\bf{y}}_{MAC}})) = {f_{NRZ}}(sign({{\bf{A}}_{MAC}}{\bf{s}}),
\end{equation}
where ${f_{NRZ}}(\cdot)$ denotes a NRZ encoder.
Here, we suppose that the security of transmitting ${\rm MAC}$ from the sender to the receiver is perfectly guaranteed. After getting the reconstructed ${\bf s}'$, the receiver can compute its ${\rm MAC}'$ and then be aware of the integrity of ${\bf s}'$ by checking the difference between the received ${\rm MAC}$ and the computed ${\rm MAC}'$. 

In light of the central limit theory, the binary sequence $MAC$ follows the Bernoulli distribution because ${{\bf y}_{MAC}}$ satisfies the sub-Gaussain distribution.
Such a $MAC$ can tolerate some energy-limited noises but some malicious manipulations.

\subsection{Access control}
In the general outsourced service, the enduser loses the ownership on cloud-stored data. Hence, the access management is an extremely important work, which ensure
the data confidentiality by restricting access only to the authorized requesters.
After receiving the request for accessing to the reconstructed data, how to distinguish legitimate access and malicious access is a significant challenge all the time. 

An access passwards generation algorithm based on the random diagonal matrix ${\bf R}_{k_3}=diag(r_1,r_2,...,r_N)$ is proposed to restrict malicious access at a very low complexity. As mentioned previously, the key-controlled ${\bf R}_{k_3}$ plays a pivotal role in the SSR service, whose diagonal entries are Rademacher random variables. Those diagonal entries cover the all information which SSR needs such that they are regarded as the shared secret between the data producer and the cloud provider. 

In this paper, those diagonal entries are used to generate the access password and then it is shared with the legitimate enduser. Let $n$ to be the length of access password. 
Here, assume that $n$ is the divisor of $N$.
The access passwords $AP$ can be generated from ${\bf r}$ as follows:
\begin{equation}
{\bf w }= \frac{1}{2}\left( {{{\bf R}_{{k_3}}} + {{\bf I}^{N \times N}}} \right) = diag({w_1},{w_2},...,{w_N}) \in \{ 0,1\} 
\end{equation}
where ${\bf I}^{N \times N}$ denotes a $N \times N$ identity matrix. The acquired ${\bf w } = [{w_1},{w_2},...,{w_N}] $ is a binary sequence. Our goal is that a $n$-bits random access passward $AP$ is extract from ${\bf w }$ and no information about ${\bf w }$ can be getted from $AP$. Let $L = {N \mathord{\left/
		{\vphantom {N n}} \right.
		\kern-\nulldelimiterspace} n}$.
Two functions $f$ and $g$ are defined as follows:
\begin{equation}
{\bf p } = f({\bf w }) = f([{w_1},{w_2},...,{w_n}])= [{p_1},{p_2},...,{p_n}]
\end{equation}
\begin{equation}
{\bf q } = f({\bf w }) = g([{w_1},{w_2},...,{w_n}])= [{q_1},{q_2},...,{q_n}]
\end{equation}
where
\begin{equation}
{p_i} = \left\{ {\begin{array}{*{20}{c}}
	{{w_1} \otimes {w_2} \otimes  \cdots  \otimes {w_L}}\\
	{{p_{i - 1}} \otimes {w_{(i - 1)L + 1}} \otimes {w_{(i - 1)L + 2}} \otimes  \cdots  \otimes {w_{iL}}}
	\end{array}} \right.\begin{array}{*{20}{c}}
{i = 1},\\
{1 < i \le n},
\end{array}
\end{equation}

\begin{equation}
{q_{n + 1 - i}} = \left\{ {\begin{array}{*{20}{c}}
	{{w_{(i - 1)L + 1}} \otimes {w_{(i - 1)L + 2}} \otimes  \cdots  \otimes {w_{nL}}}\\
	{{q_{n - i}} \otimes {w_{(i - 1)L + 1}} \otimes {w_{(i - 1)L + 2}} \otimes  \cdots  \otimes {w_{nL}}}
	\end{array}\begin{array}{*{20}{c}}
	{i = n},\\
	{1 \le i < n}.
	\end{array}} \right.
\end{equation}
the $AP$ is generated by calculating $AP= {\bf p } \otimes {\bf q } $, namely
\begin{equation}
\begin{array}{l}
AP = {\bf p } \otimes {\bf q } = [{p_1} \otimes {q_1},{p_2} \otimes {q_2},...,{p_n} \otimes {q_n}]\\
\end{array}
\end{equation}

Obviously, the access passwords are $n$ bits and could not leak ${\bf R}_{k_3}$ at least computationally. The access passwords can be generated in the sensing device and the cloud. The former needs to be secretly sent to the legitimate end user. After receiving an access request, the cloud manager asks the requester to submit their access password and then check it. If there exist any deviation, the requester is viewed as an unauthorized one so this access is rejected, ortherwise accepted.
\section{The proposed secure outsourcing framework} 
	The section detailedly introduce the proposed secure outsourcing framework from data encoding and decoding phases. Data encoding is implemented in the resource-limited sensor but data decoding in both the cloud server and the terminal device. Note that the fixed orthogonal basis matrix ${\bf \Phi}$ and hash matrix ${\bf A}_{MAC}$ are kept public. $AP$, $MAC$, and the secret keys $(k_1,k_2,k_3)$ are secretly transmitted by asymmetric cryptographic techniques or secure channels.
\begin{figure*}[t]
    \centering
	\includegraphics[scale=0.93]{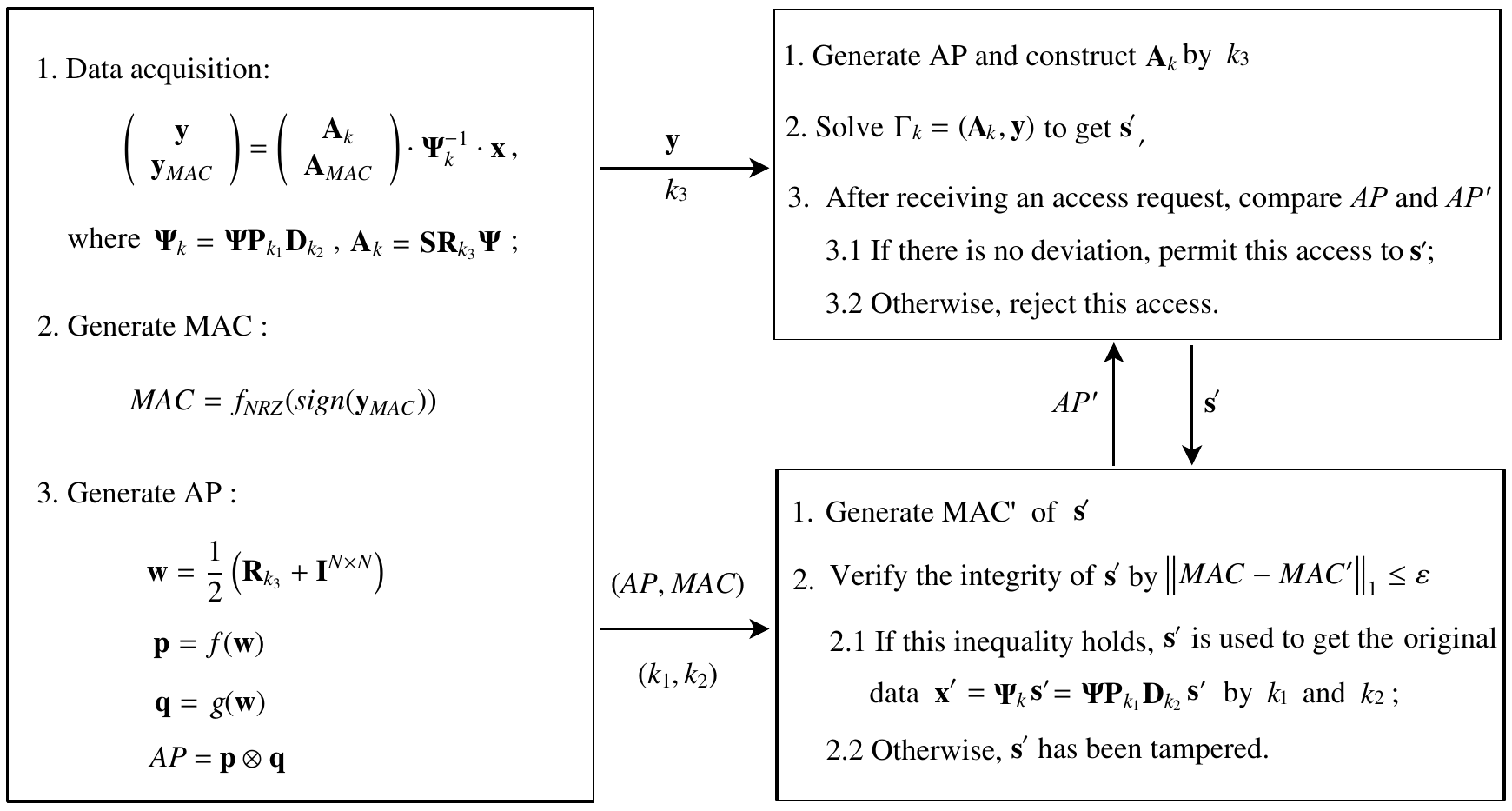} 
	\caption{Privacy-assured outsourcing framework.} 
	\label{fig:4.0}
\end{figure*}

\subsection{Data encoding phase}
There are three main operations in the encoding phase, including data acquisition, message authentication codes generation, and access passwords generation. 

\emph{1) Data acquisition:} 
\begin{enumerate}[Step 1.]
	\item Three secret keys $(k_1, k_2, k_3)$ are acquired by key agreement protocol or a secure channel.
	\item $k_1$ and $k_2$ are used to generate a random permutation matrix ${{\bf P}_{k_1}}$ and a random diagonal matrix ${{\bf D}_{k_2}}$ with non zero entries, respectively. As a result, the basis matrix can be constructed by $ {{\bf \Psi} _{k}} = {{\bf \Psi} }{{\bf P}_{k_1}}{{\bf D}_{k_2}}$.
	\item $k_3$ is used to generate a random Rademacher matrix ${\bf R}_{k_3}$. As a result, the sensing matrix ${\bf A}_{k} $ can be constructed by ${\bf A}_{k} = {\bf S}{\bf R}_{k_3} {\bf \Psi}$.
	\item The sampling-compression-encryption-hash operation is simultaneously implemented to capture $\bf y$ and ${\bf y}_{MAC}$ by the following liner projection:
	\begin{equation}
	\left( {\begin{array}{*{20}{c}}
		{\bf y}\\
		{{{\bf y}_{MAC}}}
		\end{array}} \right) = \left( {\begin{array}{*{20}{c}}
		{{{\bf A}_k}}\\
		{{{\bf A}_{MAC}}}
		\end{array}} \right) \cdot {\bf \Psi} _k^{ - 1} \cdot {\bf x}.
	\end{equation}
\end{enumerate}

\emph{2) Message authentication codes generation:} The $MAC$ of the encrypted sparse coefficients $\bf s$ are generated by extracting the sign of values as follows:
\begin{equation}
MAC =  {f_{NRZ}}(sign({{\bf y}_{MAC}})).
\end{equation}

\emph{3) Access passwords generation:}
\begin{enumerate}[Step 1.]
	\item The diagonal entries of ${\bf R}_{k_3}$ are extracted as a Rademacher random vector ${\bf r} = [r_1,r_2,...,r_N] \in \{ -1,1\} $.
	\item ${\bf r}$ is transformed as a symmetric Bernoulli random vector ${\bf w} = [w_1,w_2,...,w_N] \in \{ 0,1\}$ by ${\bf w} ={\bf r} +1$.
	\item Let $L$ to be step length and $n = \left\lfloor {{N \mathord{\left/
				{\vphantom {N L}} \right.
				\kern-\nulldelimiterspace} L}} \right\rfloor $. A $n$-bits binary sequence ${\bf p}=[p_1,p_2,...,p_n]$ is generated as follows:
	\begin{equation}
	\begin{array}{l}
	{p_1} = {w_1} \otimes {w_2} \otimes  \cdots  \otimes {w_L}\\
	{p_2} = {p_1} \otimes {w_{L + 1}} \otimes {w_{L + 2}} \otimes  \cdots  \otimes {w_{2L}}\\
	\vdots \\
	{p_n} = {p_{n - 1}} \otimes {w_{(n - 1)L + 1}} \otimes {w_{(n - 1)L + 2}} \otimes  \cdots  \otimes {w_{nL}}
	\end{array}
	\end{equation}
	\item A $n$-bits binary sequence ${\bf q}=[q_1,q_2,...,q_n]$ is generated as follows:
	\begin{equation}
	\begin{array}{l}
	{q_1} = {w_{(n - 1)L + 1}} \otimes {w_{(n - 1)L + 2}} \otimes  \cdots  \otimes {w_{nL}}\\
	{q_2} = {q_1} \otimes {w_{(n - 2)L + 1}} \otimes {w_{(n - 2)L + 2}} \otimes  \cdots  \otimes {w_{(n-1)L}}\\
	\vdots \\
	{q_n} = {q_{n - 1}} \otimes {w_{ 1}} \otimes {w_{ 2}} \otimes  \cdots  \otimes {w_{L}}
	\end{array}	
	\end{equation}
	\item A $n$-bits access passwords $AP$ can be generated by $AP={\bf p}\otimes{\bf q} = [{p_1} \otimes {q_1},{p_2} \otimes {q_2},...,{p_n} \otimes {q_n}]$. Note that $n$ should not be less than 128 in consideration of security.
\end{enumerate}	

\subsection{Data decoding phase}

\emph{1) Solving SSR problem:} 
\begin{enumerate}[Step 1.]
	\item The sensing matrix ${\bf A}_{k} = {\bf S}{\bf R}_{k_3} {\bf \Psi}$ is constructed through $k$ in the same way as the encoder does.
	\item The privacy-preserving $\bf s$ is got from the measurements $\bf y$ by solving SSR problem ${\Gamma}_k= ({\bf A}_k,{\bf y})$ in the cloud server. 
\end{enumerate}	

\emph{2) Access control:} 
\begin{enumerate}[Step 1.]
	\item The access passwords $AP$ are generated through $k_3$ in the same way as the encoder does.
	\item After receiving the access request, the correct $AP$ are compared with the submitted $AP'$. If there is no any deviation, this access is considered to be authorized, otherwise rejected.
\end{enumerate}		

\emph{3) Integrity verification:} 
\begin{enumerate}[Step 1.]
	\item After receiving the reconstructed data vector ${\bf s}'$, the end-user computes its $MAC'$.
	\item Check $\left\| {MAC - MAC'} \right\| \le \varepsilon $. If this inequality holds, ${\bf s}'$ is viewed to be acceptable, otherwise tampered. 
\end{enumerate}	

\emph{4) Recovering original data:} 
\begin{enumerate}[Step 1.]
	\item  The basis matrix $ {{\bf \Psi} _{k}} = {{\bf \Psi} }{{\bf P}_{k_1}}{{\bf D}_{k_2}}$ is constructed through $k_1$ and $k_2$ in the same way as the encoder does.
	\item The original data vector ${\bf x}'$  (or called content-preserving data) is recovered from the verified ${\bf s}'$  by computing ${\bf x}' = {{\bf \Psi} _{k}} {\bf s}'$.
\end{enumerate}	
\section{Experimental results and security analysis}
Three kinds of security vulnerabilities are considered in the proposed outsourcing architecture, including privacy leaking, data tampering, and malicious access.
In this section, we demonstrate the security of the proposed framework in terms of data confidentiality, data integrity, and access control. Simulations run in MATLAB R2015B with Core i5-7200U CPU and 4GB RAM.


\subsection{Confidentiality analysis}
Both transmission confidentiality from sensor to cloud and outsourcing computation privacy are guaranteed by the proposed bi-level encryption framework, composed of SR-based encryption and CS-based encryption. Essentially, SR-based encryption is equivalent to performing permutation-substitution operations in the transformation domain and does not change the sparsity. For more intuitive experimental results, its encrypted results are exhibited in the time domain. CS-based encryption aims at data obfuscation while randomly subsampling. Such a product cipher is perfectly built in the sensing player at almost zero cost and provides the captured data with innate confidentiality. Here, we investigate its privacy-preserving capability by simulation experiments and theoretical analysis below.

\begin{figure*}[t]
	\centering
	\subfigure[Plain images]{
		\includegraphics[scale=0.832]{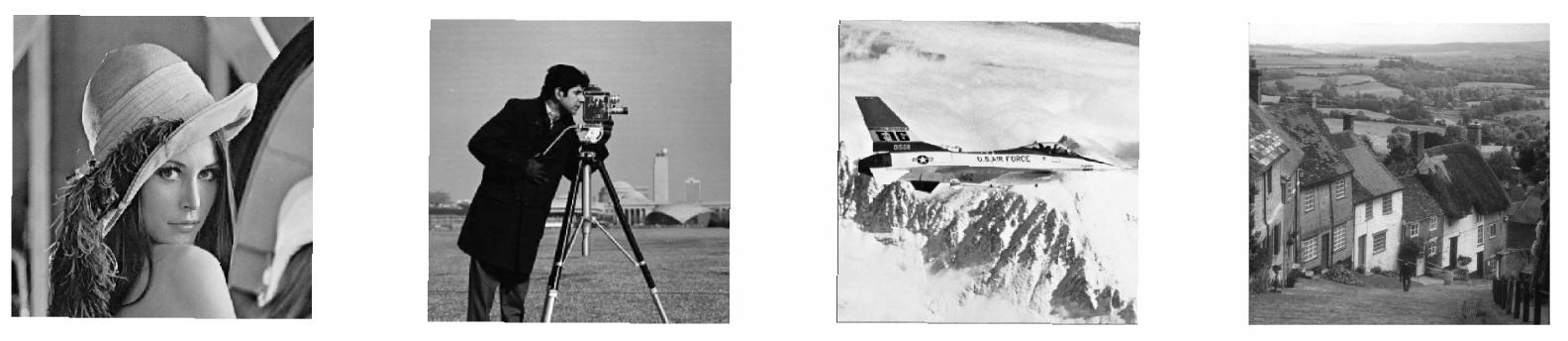} 
	}
	\subfigure[Halfway encrypted images]{
		\includegraphics[scale=0.832]{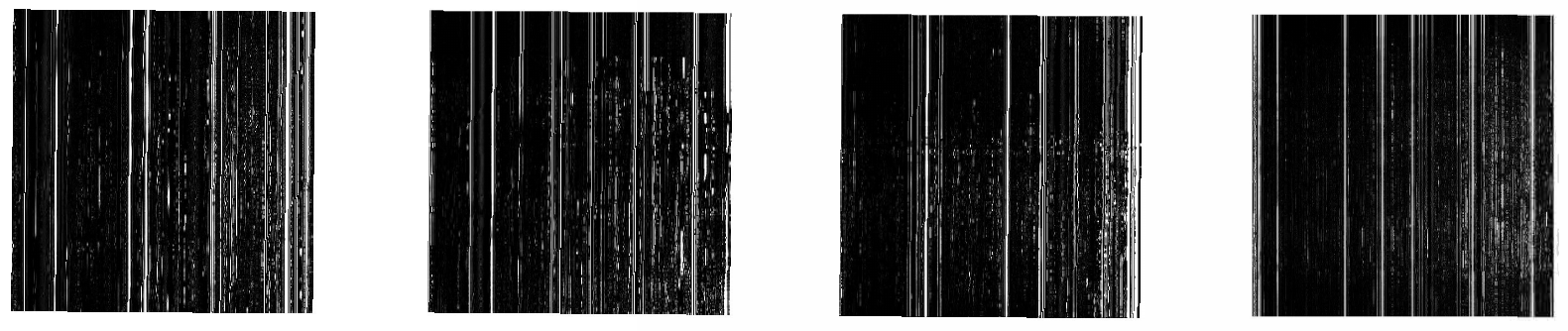} 
	}
	\subfigure[Finally encrypted images]{
		\includegraphics[scale=0.832]{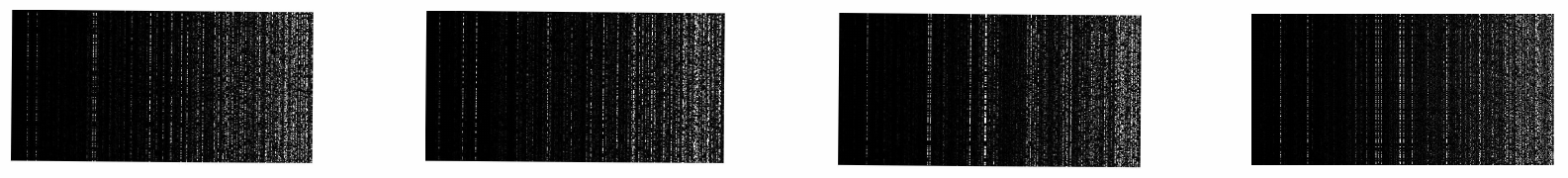} 
	}
	\subfigure[Decrypted images]{
	\includegraphics[scale=0.832]{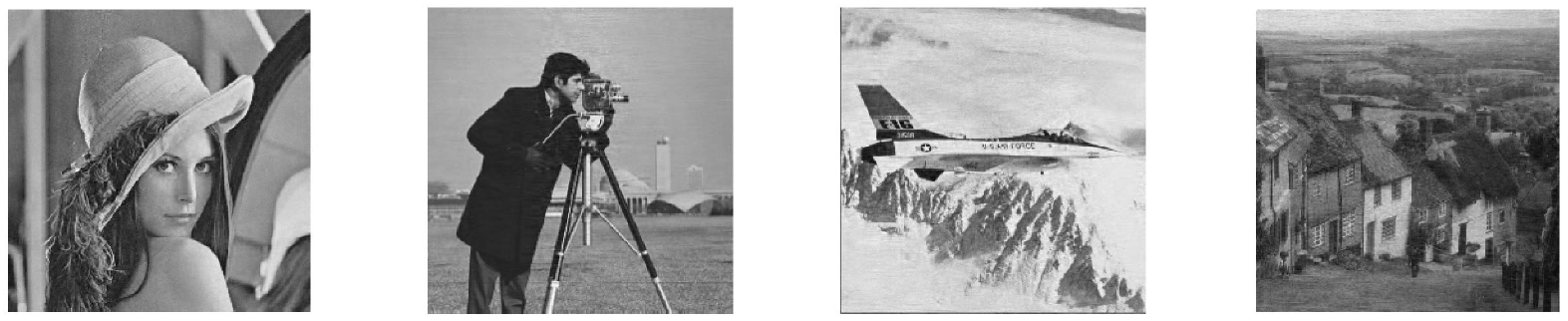} 
}
	\caption{Encrypted and decrypted results for four plain images in $CR=0.5$.} 
	\label{fig6.1}
\end{figure*}

Compressive ratio $CR$ is set to 0.5. The four plain images are encrypted through SR, then encrypted through CS, and finally decrypted as shown in Fig. \ref{fig6.1}. The finally encrypted image containing less data are transmitted from the sensor and the cloud. The halfway encrypted image contains all available information for the cloud. Obviously, both the halfway encrypted images and the finally encrypted images do not leak any visually meaningful information. Note that the halfway encrypted image leaks the sparsity of the original signal. The visual secrecy provided by SR-based encryption is acceptable in the face of the semi-trusted cloud but not enough to guarantee secure transmission over the public channels. Such a vulnerability occurred in \cite{zhang2015support,hu2017compressive,zhang2019efficiently}. Next, we consider some common attack scenarios in the process of transmission, including brute-force attack, ciphertext-only attack (COA), and plaintext attack.
 
\emph{1) Brute-force attack:} As mentioned previously, both the sensing matrix and the basis matrix are controlled by several keys, namely $ {{\bf \Psi} _{k}} = {{\bf \Psi} }{{\bf P}_{k_1}}{{\bf D}_{k_2}}$, ${\bf A}_{k} = {\bf S}{\bf R}_{k_3} {\bf \Psi}$. The whole encoding process depends on three random matrix $({{\bf P}_{k_1}}, {{\bf D}_{k_2}}, {\bf R}_{k_3})$. Hence, the potential case of brute-force attack is that the attacker tries to get the real matrices $({{\bf P}_{k_1}}, {{\bf D}_{k_2}}, {\bf R}_{k_3})$ by guessing or searching the whole key space of $(k_1,k_2,k_3)$ exhaustively and then uses them for decoding.

 Let $Pro(\cdot)$ denotes the probability of a successful trial. It is not hard to calculate $Pro({{\bf P}_{k_1}}) = {1 \mathord{\left/
 		{\vphantom {1 N}} \right.
 		\kern-\nulldelimiterspace} N}!$, $Pro({{\bf D}_{k_2}})  = {1 \mathord{\left/
 		{\vphantom {1 {{2^N}}}} \right.
 		\kern-\nulldelimiterspace} {{f^N}}}$, $Pro({{\bf R}_{k_3}})  = {1 \mathord{\left/
 		{\vphantom {1 {{2^N}}}} \right.
 		\kern-\nulldelimiterspace} {{2^N}}}$. So the probability of successful guessing the three matrices is $Pro({{\bf P}_{k_1}}, {{\bf D}_{k_2}}, {\bf R}_{k_3}) = {1 \mathord{\left/
 		{\vphantom {1 {(N! \cdot {f^N} \cdot {2^N}}}} \right.
 		\kern-\nulldelimiterspace} {(N! \cdot {f^N} \cdot {2^N}}})$. For a $N$-conditionality multimedia data, $N$ is often enough large to make this brute-force attack infeasible, namely $Pro({{\bf P}_{k_1}}, {{\bf D}_{k_2}}, {\bf R}_{k_3})\\ \approx 0$.

\emph{2) Cipher text-only attack:} A recent work pointed out that CS measurements acquired by a structurally subsampled sensing matrix are said to be indistinguishable as long as the original signal has constant energy\cite{cho2019secure}. Such a CS-based cryptosystem can achieve the asymptotic spherical secrecy\cite{Cambareri2015Low}. It means that $\bf y$ only leak the energy information of $\bf x$ when a structurally subsampled sensing matrix is employed, rooted in the energy-preserving constraint of RIP condition.
However, a non-normalized random matrix ${{\bf D}_{k_2}}$ is imported to CS-based encoding in the proposed framework. It can break the energy conservation between $\bf x$ and $\bf y$ but does not affect the exact reconstruction such that a higher secrecy than asymptotic spherical secrecy is achieved in the proposed framework.

\begin{theorem}
\cite{cho2019secure}	A CS-based cryptosystem ${\bf y} = {\bf A} {\bf x}$ can be said to achieve asymptotic spherical secrecy as long a $\bf A$ is a structurally subsampled sensing matrix, where $\bf x$ and $\bf y$ denote the spare plaintext and the corresponding ciphertext, respectively.
\end{theorem}
\begin{definition}
	(Asymptotic spherical secrecy\cite{Cambareri2015Low}) Define ${\bf x} = [{x_1},{x_2},{x_3},...,{x_n}]$ as a plaintext sequence and ${\bf y}$ as the corresponding ciphertext sequence. Assume that the energy of the plaintext ${\varepsilon_{\bf x}}$ is finite. Such a cryptosystem is said to achieve asymptotic spherical secrecy if ${f_{{\bf y}|{\bf x}}}(x,y)\mathop  \to \limits_{\mathcal D} {f_{{\bf y}|{\varepsilon_{\bf x}}}}(y)$, where ${\mathcal D}$ denotes the convergence in the distribution as $n \to \infty$.
\end{definition}

\emph{2) Plaintext attack:}  
From the perspective of stealing the complete plaintext, It is widely accepted that CS-based basic cryptosystem is immune to brute-force attack and COA. Special attention is paid to the plaintext attacks. Plaintext attack means that abundant plaintext-ciphertext pairs have been captured. When the measurement matrix $\bf \Phi$ is fixed, the attacker can figure out $bf \Phi$ from $N$ independent plaintext-ciphertext pairs and further illegally recover the current plaintext by leveraging $\bf \Phi$. To resist plaintext attack, basic CS cipher is forced to renew $\bf \Phi$ frequently, even in a one-time sensing (OTS) model.

In the proposed framework, the measurement matrix is $\bf \Phi =  {\bf S}{\bf R}_{k_3} {\bf \Psi} ({{\bf \Psi} }{{\bf P}_{k_1}}{{\bf D}_{k_2}})^{ - 1}$. $ {\bf S}{\bf R}_{k_3}{\bf \Psi}$ has been demonstrated to be a RIP matrix. ${{\bf D}_{k_2}}$ breaks the energy-preserving guarantee of RIP such that $\bf \Phi$ is a non-RIP matrix. In a multi-time sensing (MTS) model, $\bf \Phi$ can also be calculated from $N$ independent plaintext-ciphertext pairs. The difficulty of plaintext attack depends on the complexity of getting $\bf D$ by decomposing $\bf \Phi$ or guessing. No matter which approaches, it seems to be an impossible task in polynomial time.

Therefore, without updating the key $(k_1,k_2,k_3)$ frequently, the proposed framework can provide outsourced data with a high level of confidentiality against brute-force attack, COA, and plaintext attack.  

\subsection{Integrity analysis}

A noise-robust but tamper-resistant integrity verification approach is involved in the proposed framework. Considering that both transmission and outsourcing computations carry out in the ciphertext domain, the target of tampering the plaintext data $\bf x$ 
seems to be achieved only in two cases. One that polluting measurements $\bf y$ makes the reconstruction process fail or the reconstructed data beyond recognition. The other is that $\bf y$ is replaced with anther fake containing entirely different information, based on the worst scenario where the eavesdropper has grasped the practical measurement matrix $\bf \Phi$. Therefore, we discuss four situations as described in Fig. 4, namely no noise, normal noise, malicious pollution, and fake image. It is no surprise that $MAC$ generated from the recovered data are different from one generated from the original data. The bit error rate (BER) of $MAC$ is used to estimate the distortion degree of $MAC$. The experimental results are shown in Fig. 5. There is almost no change in $MAC$ under no noise circumstance. The normal noise makes a slight difference to $MAC$ but malicious noise dose the opposite. The fake data can also change $MAC$ to some extent. Obviously, data pollution and replacement behaviors can be easily detected when the threshold value of BER is set in the interval $(0.1, 0.15)$. The performance of such a data integrity verification approach is affected by CR, noise, reconstruction algorithm, etc. Note that some small-region data tamper may be viewed as the normal noise.
\begin{figure*}[t]
	\centering
	\subfigure[]{
		\includegraphics[scale=0.22]{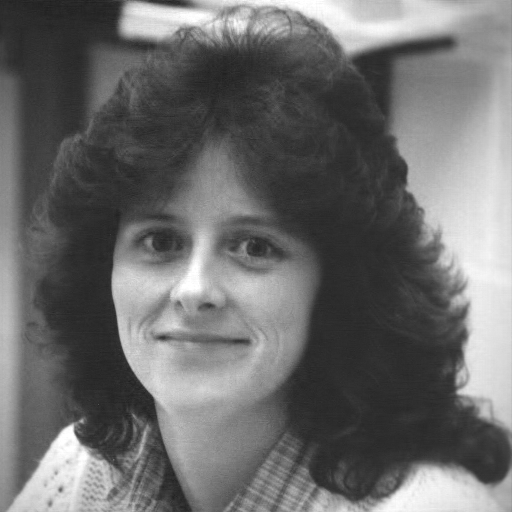}
	}
	\subfigure[]{
		\includegraphics[scale=0.22]{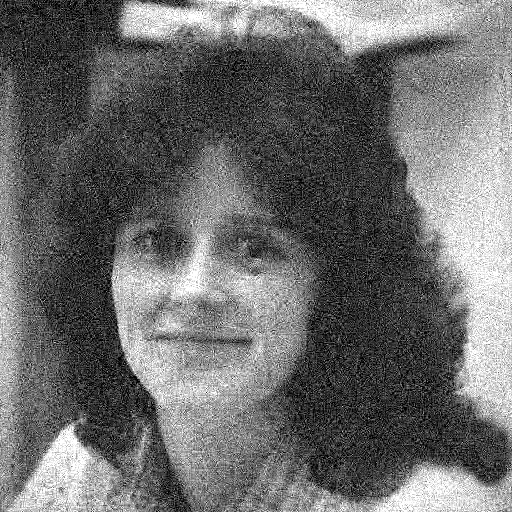}
	}
	\subfigure[]{
		\includegraphics[scale=0.22]{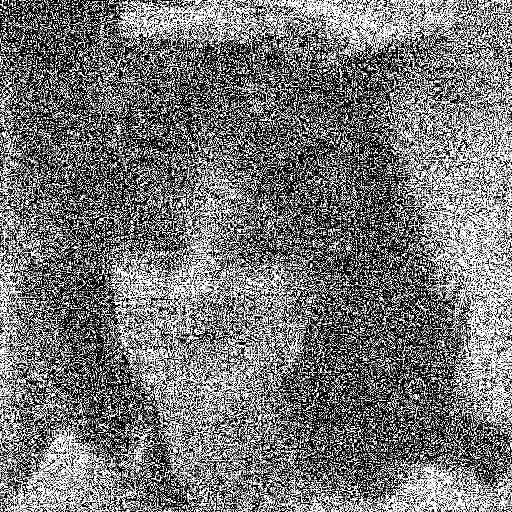}
	}
	\subfigure[]{
		\includegraphics[scale=0.22]{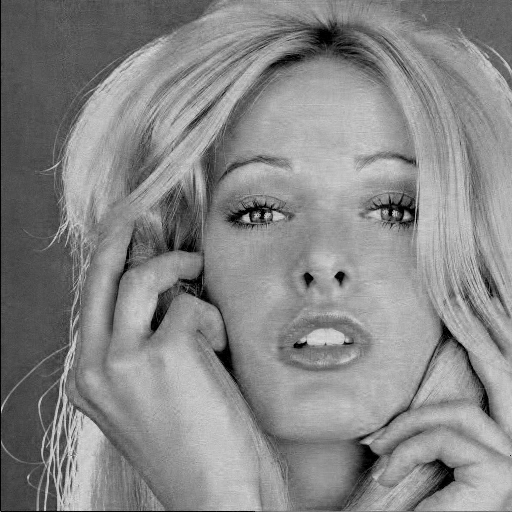}
	}
	\caption{The reconstructed results in four different situations. (a) no noise; (b) normal noise; (c) malicious pollution; (c) fake image.}
	\label{fig_6}
\end{figure*}
\begin{figure}[ht]
	\centering
	\includegraphics[scale=0.55]{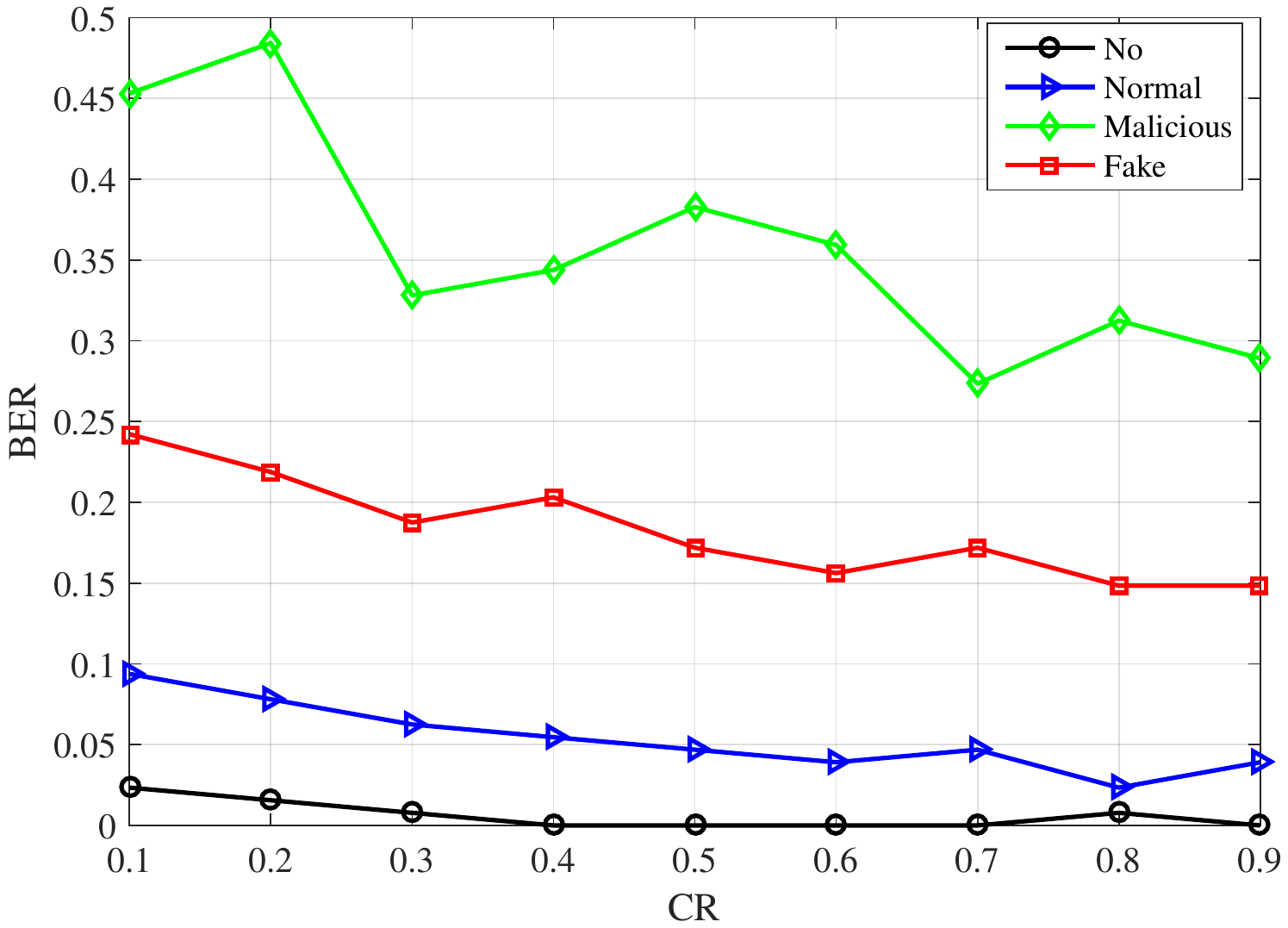} 
	\caption{BER versus CR in different situations.} 
	\label{fig:5.0}
\end{figure}

\subsection{Access control analysis}	
To restrict the malicious access, a lightweight access passwords $AP$ generation algorithm is proposed in Section 4, which is based on the prior knowledge about sensing matrix. The end-user launches an access request by submitting $AP$ and the cloud manager verifies the validity of this access by checking $AP$. The transmission security of $AP$ depends on the key agreement protocols, which is outside the scope of this paper. Taking security into consideration, we suggest that $AP$ should be at least 128-bits. Here, we test the sensitivity of access passwords generation algorithm. Essentially, the proposed algorithm is to transform a length-variable Bernoulli random sequence to an unpredictable 128-bits binary sequence. Three kinds of data manipulations are performed on the input sequence, including inverting some elements, randomly permuting two elements, slightly key perturbation. As shown in Table 1, the output AP would be radically changed no matter which manipulation is performed. It means that each encoding would generate an exclusive access password with high probability. Only by submitting a correct access password can the access requester be considered as an authorized user.

\begin{table*}[t]
	\centering
	\caption{Access password test}\label{tab:7}
	\begin{tabular}{ p{2cm}<{\centering}  p{2cm}<{\centering}  p{2cm}<{\centering}  p{2cm}<{\centering}  p{2cm}<{\centering}  p{2cm}<{\centering}  p{2cm}<{\centering}}
		\toprule
		Manipulation types &   & Inverse first element &Inverse middle element& Inverse last element & Permutate two elements& Key perturbation\\
		\midrule
		Access passwords &  
		6BF27A9B
		B956A7BA
		5FDF81DB
		C394F691   & 
		 21DFF11E
		CF2EC346
		85EE6CA9
		A841AA7E & 
		38D0BFD1
		109A1BF2
		9D09D896
		CE0340C6 & 
		D3D1AF9A
		6BE27A9B
		B956B7BA
		7F4E83CB  & 
		E2EA6787
		89AAFA71
		06D72E17
		A12C67DC  & 
		725E7524
		3C61D8ED
		F46B4103
		09036C4A \\
		Change rate &    &  95.14\%  &  97.73\% & 95.62\%  & 94.35\% &  96.81\%  \\
	
		\bottomrule              
	\end{tabular}
\end{table*}


\section{Conclusion}
This paper considers such a scenario in which both sensing device and terminal device are resource-constrained but the cloud is resource-abundant, which generally appears in the current IoT applications. With the boom of multimedia big data, it is an predictable trend to transform computational complexity form the local side to the cloud side. In this paper, a privacy-assured outsourcing scheme is proposed for multimedia, in which CS plays a pivotal role. CS's low-complexity encoding is carried forward in the resource-constrained sensing device and high-complexity decoding is left to the cloud. More importantly, several security issues are effectively solved in CS framework.

A CS-based product cipher is proposed to provide the outsourced data with secrecy guarantee in the process of transmission and computation. Security analyses indicate that the captured data are able to resist brute-force attack, COA, and plaintext attack. Considering the malicious access attack to cloud, an access password generation algorithm is proposed, which is based on the prior knowledge of the structurally subsampled sensing matrix. In addition, data integrity guarantee is also acquired by a CS-based message authentication codes generator. Differing from most sensitive hash algorithm, CS-based message authentication codes can not only resist noise and tamper attacks but tolerate energy-finite noise to some extent.

\bibliographystyle{elsarticle-num}
\bibliography{ref}
\end{document}